\documentclass[journal]{IEEEtran}

\usepackage{booktabs}
\usepackage{array}
\usepackage{graphicx}
\usepackage{upquote}
\usepackage[hyphens]{url}
\usepackage[hidelinks]{hyperref}

\begin{document}

\title{The Compiler May Read It, the Agent May Not: Keeping Part of a Research
Code Away from a Coding Agent}

\author{Shobhan~Roy%
\thanks{S. Roy is with the University of Iowa, Iowa City, IA 52242 USA
(e-mail: shobhan-roy@uiowa.edu).}%
\thanks{Submitted to IEEE Computing in Science \& Engineering. This work has been submitted to the IEEE for possible publication. Copyright may be transferred without notice, after which this version may no longer be accessible.}}

\markboth{Roy: The Compiler May Read It, the Agent May Not}{Roy: The Compiler May Read It, the Agent May Not}

\maketitle

\begin{abstract}
The compiler must read modules a physics-based solver cannot build without; the coding agent must not read that intellectual property. The harness, the software around the model that executes its tool calls, does
not ship that rule. We classified fifteen read routes against a container,
permission rules and a sandbox. None of the three can tell which program
is reading.
\end{abstract}


\section{Introduction}
\IEEEPARstart{W}{e} run a coding agent on a physics-based numerical code, a flow solver for the
partial differential equations of fluid motion. The agent edits the solver,
builds it and reads the compiler's diagnostics to correct its edits. This
edit-build-fix loop is the reason to put an agent on a research code.
Solvers of this kind are commonly built on an open-source base, which supplies
the discretization and the parallel machinery, and completed by a group's
own material and reaction models. Those models are the constitutive and kinetic
closures that make the governing equations solvable. They can hold a group's
intellectual property ahead of publication, and some are licensed to the group
under terms that allow generating publishable data but not passing on the
algorithms themselves. If the agent reads the files that hold them, their
contents go into a transcript that leaves the machine. The solver does not
build without those files. Consequently the compiler has to open them. The rule
we wanted, which is simple to state, is that the compiler may read those files
but the agent may not. Four
mechanisms stand between the agent and the files: a) the container the agent
runs in dictates the runtime user and the paths mounted for the agent; b) the
harness's permission rules match on the text of a command and on a path; c) the
harness's sandbox matches on a path; d) the plain-language instruction file the
agent reads when each session starts is a soft contract that can state the rule
in full but does not enforce it. As shipped, none of the three that enforce anything can state the criterion
the rule turns on, namely which program is reading.

The article makes four observations. The first is what each of the four
mechanisms can and cannot say about a read. The second is a table of fifteen
ways the agent can reach a protected file, classified against the configuration
we run. In that configuration, the permission rules and the sandbox cover different routes, and
five routes are covered by neither. The third is the history of one deny list
over five weeks. Its four revisions, made in the course of ordinary use, added
118, 13, 4 and finally zero new command names. The fourth is what a mechanism
would have to do to enforce the rule, and what it costs to approximate it with
the pieces an ordinary user already has.

\section{Terms and Trusted Components}

At the file itself, the compiler's read and the agent's read look the same to
all four mechanisms. The agent and the build run as one user account, and the
compiler is a child of the agent's own shell, inside the agent's container.
The two reads differ only in why the file is opened and where the bytes go. The
compiler's read ends on the same machine, in object code that the agent
executes. The agent's read ends in its context window, which is tokenized
and leaves the machine.

The harness is the program that runs the language model and executes its tool
calls. Its permission rules include a deny list of command and path patterns.
The harness refuses any tool call the agent requests that matches one of them.
Its sandbox is an operating-system boundary that the harness places around the
shell commands it runs.

We use two terms in a specific sense. By \emph{the agent} we mean everything the
harness drives: a) the model's context; b) the agent's own file-reading tools;
c) the shell the harness starts and every process descended from that shell; d) the hook
scripts the harness runs on the agent's behalf; e) any policy server it is
configured to call. All of them can put text in front of the model. By
\emph{may not read} we mean that the bytes of a protected file must not reach
the model's context or the transcript through any of them. We allow one channel only, the output of an ordinary build, for the sake of
operating the solver. A compiler diagnostic may quote the
line it rejects, and we accept that. We do not accept the compiler's other
output modes, which can print a file in its entirety. Written out, the rule is
that the compiler may open the protected files, that its object code and its
diagnostics may reach the agent, and that the files themselves may not, by any
path. Both permitted outputs still carry content derived from the file. The
diagnostic carries the line it quotes, and the object code carries the models'
constants. The rule is therefore a bounded disclosure and not
\emph{noninterference}, the stronger property under which nothing the agent
sees would depend on the file.

We take the position of an ordinary user of the agent. Such a user cannot
change the harness, cannot replace the language runtime, and cannot load a
kernel security policy, which lives in the host kernel that every container
shares. Accordingly every negative result in this article is relative
to that position. The negative results do not bind an administrator with root on
the host machine, who can install a kernel-level enforcer. This article examines
what these mechanisms can express and cover, and what they cannot. It does not
examine how they hold against a designated adversary.

We do not treat the model or the harness as hostile. However, the boundary
cannot rest on the agent's cooperation either. A tool call can follow from a
misreading of the task, from a misfired model output, or from text the agent
reads in a file and acts on. The parts that have to be trusted are the kernel
and the container runtime, the vendor's harness and the model service behind it,
the compiler, and any hook or policy server we install ourselves.

The build files have to be trusted as well, because a build recipe the agent can
edit, or pass flags to, can print a protected file. That trust holds if the
recipe and its flags are out of the agent's reach. In this deployment the instruction file can say so, but it enforces
nothing. That is why a recipe under an allowed build target is among the routes
no mechanism covers.

\section{The Rule in the Security Literature}

The security literature calls this rule \emph{purpose-based access control}. It
permits or refuses an access according to what the access is for. The term was
presented by Ji-Won Byun and Ninghui Li \cite{byun2008purpose}, who listed three
ways a system can learn the purpose of an access. The first is to ask the
requester, which ``requires complete trust on the users''. The second is to
register an application with a fixed purpose, which ``cannot be used for complex
stored-procedures or applications as they may access various data for multiple
purposes''. The third is to infer the purpose, which the authors judged hard to
do accurately. The instruction file is the first way, and
exempting the build from the deny list is the second. A build is essentially
such a multi-purpose application. Byun and Li's remedy was to check a declared
purpose against something the requester does not control. However, a vendor-run harness offers nothing of the kind, because the agent
composes the command line and every attribute the permission rules inspect is
agent-authored. We use purpose
in these authors' sense, the use the requester intends. No operating system
observes it directly. A kernel can observe only stand-ins for it, namely which
process opened the file, what that process descends from, and where its output
goes. This article's finding is therefore about the stand-ins rather than about
purpose itself. The finding is that none of the three
mechanisms that enforce anything can name any of the stand-ins.

The idea of deciding a read by which program is doing the reading has been
documented before, under the name \emph{type enforcement}. W. Earl Boebert and
Richard Y. Kain put each running program in a domain and each file in a type,
and wrote down in a table what every domain may do to every type
\cite{boebert1985type}. An entry left out of the table was a refusal, and
because the table was finite and explicit an auditor could check the whole
policy by reading it. A list of forbidden command names has neither property.
It permits whatever it omits, and, at the length such lists reach, it can no
longer be checked by reading. Linux carries type enforcement in the kernel as
SELinux, which moves a process into a new domain when it executes a program of a
given type. If the protected files are given a type that the agent's domain
cannot read, the kernel refuses the agent while the compiler proceeds. That needs a privileged policy loader for a kernel we do not administer.

The classical alternative, a \emph{trusted broker}, needs none of that, and an
ordinary user can deploy it. The protected files go under a second user account,
outside the agent's filesystem, and the agent reaches the build only through a
channel that account controls: a) a build recipe the agent cannot edit; b) a
request that names a revision of the agent's code and nothing else; c) a reply
filtered down to what the agent needs in order to act.

The broker approximates the rule rather than enforcing it, and the
approximation has two costs. The first is that the build's own output carries
bytes from the protected files. A compiler diagnostic quotes the line it
rejects, and preprocessed source can carry the whole file. The filter on the
reply is therefore the actual security control. Widening it lets more of the
protected file back through. Our own arrangement already concedes the
diagnostic, since it lets the agent read one while refusing it the file. The
second cost is the edit-build-fix loop. A broker that answers only pass or fail
removes that loop. One that returns the failure message hands back the
offending line. In addition, someone other than the agent has to own the
recipe, so build engineering moves back to a person. The broker is the right
choice where the protected files are stable and separate from the code the agent
edits, but where they are themselves under development it removes the loop the
agent was brought in for. Our result is therefore narrow. The controls this
harness exposes cannot express the rule, and a separately trusted mediator
outside the harness can only approximate it.

\section{The Blind Spot of Matching on Names}

\begin{table*}[t]
\caption{Route coverage. Fifteen ways the protected bytes can reach the agent's
context, the command for each (\texttt{f} is the protected file), and which
mechanism covers the route in this deployment. Sources, as
superscripts: (C) a 2003 catalogue of indirect paths to a resource, (V) the
vendor's own statements of scope, (A) shell constructs from the author's earlier
published disclosure, (O) a published demonstration with the same agent, (D)
follows from the shipped rule set and the documented matching semantics. The
first two routes need no source.}
\label{tab:coverage}
\centering
\footnotesize
\setlength{\tabcolsep}{5pt}
\resizebox{\textwidth}{!}{%
\begin{tabular}{@{}llcccc@{}}
\toprule
route & command & container & permission & sandbox & instruction \\
\emph{what it names} & & \emph{account,} & \emph{command text,} & \emph{a path} & \emph{a purpose} \\
 & & \emph{mounts} & \emph{a path} & & \\
\midrule
direct read                                     & \texttt{cat f}                                         & $\circ$ & $\bullet$ & $\bullet$ & -- \\
byte-to-text encoder                            & \texttt{base64 f}                                      & $\circ$ & $\bullet$ & $\bullet$ & -- \\
interpreter, path in a literal$^{\mathrm{A}}$   & \texttt{python3 -c "open('f').read()"}                 & $\circ$ & $\circ$   & $\bullet$ & -- \\
shell builtin redirection$^{\mathrm{A}}$        & \texttt{while IFS= read -r l; do echo "\$l"; done < f}  & $\circ$ & $\circ$   & $\bullet$ & -- \\
file-descriptor indirection$^{\mathrm{C}}$      & \texttt{exec 3<f; cat <\&3}                            & $\circ$ & $\circ$   & $\bullet$ & -- \\
symlink redirection$^{\mathrm{C}}$              & \texttt{ln -s f /tmp/x; cat /tmp/x}                    & $\circ$ & $\circ$   & $\bullet$ & -- \\
process substitution$^{\mathrm{A}}$             & \texttt{grep . <(cat f)}                               & $\circ$ & $\circ$   & $\bullet$ & -- \\
another path to the same binary$^{\mathrm{C}}$  & \texttt{/proc/self/root/usr/bin/cat f}                 & $\circ$ & $\circ$   & $\bullet$ & -- \\
binary loaded by the dynamic linker$^{\mathrm{O}}$ & \texttt{/lib64/ld-linux-x86-64.so.2 /bin/cat f}     & $\circ$ & $\circ$   & $\bullet$ & -- \\
\addlinespace[1pt]
the compiler in an output mode$^{\mathrm{D}}$   & \texttt{gfortran -E f}                                 & $\circ$ & $\circ$   & $\circ$   & -- \\
recipe under an allowed target$^{\mathrm{D}}$   & \texttt{make dump}, whose recipe runs \texttt{cat f}   & $\circ$ & $\circ$   & $\circ$   & -- \\
version-control history read$^{\mathrm{D}}$ & \texttt{git show HEAD:f}                               & $\circ$ & $\circ$   & $\circ$   & -- \\
built-in file-read tool$^{\mathrm{V}}$          & \texttt{Read(f)}                                       & $\circ$ & $\bullet$ & $\circ$   & -- \\
policy-server file read$^{\mathrm{V}}$          & \texttt{read\_file(f)} on an MCP server                & $\circ$ & $\circ$   & $\circ$   & -- \\
hook-invoked read$^{\mathrm{V}}$                & a hook script reads \texttt{f}                         & $\circ$ & $\circ$   & $\circ$   & -- \\
\bottomrule
\end{tabular}}

\vspace{2pt}
{\footnotesize $\bullet$ covered \quad $\circ$ uncovered \quad -- states the rule, doesn't enforce it.}
\end{table*}

The documentation of the vendor we use states that the permission rules and the sandbox
do not cover the same operations. The sandbox's limits ``apply to Bash, PowerShell, and Monitor commands and the
processes they start,'' while the built-in file tools, the agent's own read and
edit operations, ``follow permission rules instead'' \cite{anthropicsandboxing}. Therefore a refusal through one of them does not establish
that the file is out of reach through the other. Splitting the check across two mechanisms is not a defect in itself, so long as the checks compose and every access meets one of them. In this deployment they do not compose. On
some routes no check runs, which is a failure of \emph{complete mediation}, the
principle that every access to every object is checked
\cite{saltzer1975protection}.

Table~\ref{tab:coverage} classifies fifteen prospective ways of getting the
protected bytes into the agent's context against the configuration we run, and
gives the command for each. Two of them are direct reads that any deny
list would name. Three are taken from a 2003 catalogue of indirect paths to a
resource \cite{garfinkel2003traps}. Three others come from the vendor's own
statements of what its sandbox does not cover \cite{anthropicsandboxing}, while three are shell constructs from the author's earlier published disclosure. One was
shown publicly with this same agent, by invoking the dynamic linker on a binary
that an enforcement rule named \cite{ona2026denylistescape}. The remaining three are consequences of the rule set itself. One is a build target the rules allow, whose recipe reads the file. Another is the compiler, which the rules must allow, invoked in an output mode that prints the file instead of compiling it. The third is the repository's own history. The protected files are under version control and the repository sits in the
agent's tree, so the version-control tool can print any of them at a revision, addressed by a path relative to the repository that no rule names.

A route is counted as covered when a rule of that mechanism applies to it in
this deployment and refuses it. The 60 cells rest on four kinds of evidence, ordered from strongest to weakest:
\begin{enumerate}
\item four cells are read directly from the shipped rule set, which names a
  command and a path;
\item five cells restate the vendor's documented scope verbatim;
\item 21 cells follow from those statements or from shell semantics,
  which is our reasoning rather than a measurement;
\item the remaining 30 are uniform by construction: the container column
  because every protected file sits inside the mounted project tree, and the
  instruction column because an instruction file enforces nothing.
\end{enumerate}
Two of the routes we have also seen refused in practice, a direct read and an
encoder invocation, both denied on the first attempt against a decoy tree in
July 2026. We did not run the other thirteen, against the protected files or the decoy. A
success against the protected files would have been a disclosure, and a failure
against either would speak to one spelling of one command rather than to the
class it stands for. The hook and the policy server are both configured in this deployment. They run
under the same user account as the agent, and the vendor's documentation places
neither of them inside the sandbox. The rules, the routes and every cell are the author's, without a second rater. The table says what each mechanism can name
in one deployment. It is not a measure of how often an attack would be stopped,
and each row stands for a class of routes rather than a count of them.

\begin{figure}[t]
\centering
\includegraphics[width=\columnwidth]{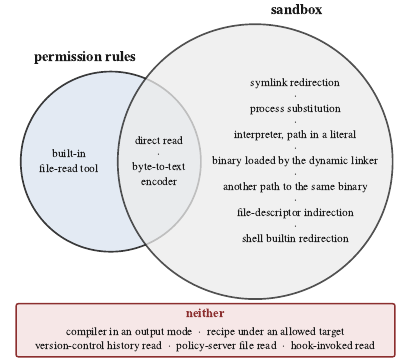}
\caption{Which mechanism covers which route. The fifteen routes of Table~\ref{tab:coverage}, placed by the mechanism whose rule refuses them in
this deployment. The permission rules and the sandbox overlap on two routes,
and five routes fall outside both.}
\label{fig:venn}
\end{figure}

Neither the permission rules nor the sandbox covers everything the other covers. As
Figure~\ref{fig:venn} shows, two routes are covered by both, seven by the
sandbox alone, one by the permission rules alone, and five by neither. Hence neither
mechanism suffices on its own, and their union cannot be expressed as one policy
through the interfaces this harness gives us. The one route only the permission
rules cover is the agent's built-in file-read tool, which the sandbox's
documented scope leaves out. The seven the sandbox alone covers evade the command match in one of three ways.
Either the read presents no command token to match, or the token it presents
is not the protected path, or the program it names is not the name on the
list.
The five that neither covers are a) a recipe under a build target the agent is
allowed to run; b) a read performed by a policy server the agent talks to over
the Model Context Protocol (MCP); c) a read performed by a hook, which is a script the
harness runs on the agent's behalf at fixed points; d) the compiler itself, asked to preprocess a protected file rather than to compile it; and e) the version-control tool, asked for the file at a revision rather than at its path. Each is a
program the agent causes to run, with the agent's authority, through a name the
rules have already permitted. The compiler route matters most. A direct read of a protected file \texttt{f} is refused while \texttt{gfortran -E f}
is not, though the compiler names the path as plainly as \texttt{cat} does.
The build cannot run unless the compiler is permitted to read the file, and the
route uses that permission. The sandbox can grant it only by exempting the
build's commands by name, whatever output mode the compiler is then asked for.

We kept one deny list under revision for five weeks, with the protection target held still at two paths. Four dated revisions took the rule count from 14 to 488, 540, 556 and 586. The first three added 118, 13 and 4 new command names, taking the distinct names from 10 to 128, 141 and 145. The fourth added 30 rules and no new name; this is the informative one. Fifteen commands already on the list had been guarded on the directory but not on the single file, and the revision extended them to the file. None of the fifteen commands opened a new path to the file's contents. The two targets had separate command lists inside the script that generated the
deny list. Nothing checked that every reader in the directory list was also in
the file list. Each rule names one command and one target, so the rule language
cannot say that two targets must be protected alike.

\section{What Enforcement Would Require}

None of the mechanisms this harness exposes names the consumer of a read. They
name a user account, a path, or the spelling of a command, and the one artifact
that does name a purpose, the instruction file, enforces nothing. Research systems do name the consumer, by process lineage \cite{actplane2026} or by a bounded access episode \cite{portico2026}, but each asks for a runtime, a kernel or an execution environment that an ordinary user does not own. This deployment has no enforcement point for the requirement.

A check that could enforce the rule would be made when a file is opened, would
know which program will consume the bytes, and would have four properties:
\begin{enumerate}
\item it runs at the open and not when a command is launched, and so needs an
  enforcement point below the tool surface, because a command names a program
  and only the open names a file;
\item it does not pass down to programs the reader starts, and so needs a way of
  tagging processes that survives the creation of new ones, because a child that
  inherits its parent's permission is the \emph{confused deputy}
  \cite{hardy1988confused}, the old problem in which a trusted program does on an
  untrusted one's behalf what the untrusted one may not do itself;
\item it cannot be satisfied by whichever process starts the consumer, since
  otherwise the agent grants itself the exemption by starting the compiler, and
  nothing we surveyed has this property;
\item it bounds what the consumer may hand back, because a compiler the agent
  can invoke, and whose output the agent can read, is a reader by proxy, the route in the compiler row of Table~\ref{tab:coverage}.
\end{enumerate}

The fourth property places the requirement beyond file-open authorization alone. Which
bytes of a protected file may appear in a compiler's output is a question of
information release rather than of who may open a file. Diagnostics are the clearest case. Compiled artifacts are another case, and they are conceded for the same reason, since the executable is what the build is for. Any compiled artifact, the executable and the compiler's intermediate files
alike, carries the numerical constants of the models it implements. It also carries
the routine names and, if built with debugging information, file paths and line
tables. Building without debugging information and stripping the symbols
narrows that channel to the constants. That is the usual position of a group that distributes a binary of a proprietary model, though it is not a guarantee against reverse engineering. Consequently the answer needs a
filter on the build's reply rather than a rule on the file. We name that
property here because the rule is incomplete without it, and we do not offer a
way to enforce it.

Other products implement part of this check, and their documentation
states where it stops. GitHub's content exclusion keeps named files out of Copilot's completions
and chat, and GitHub records that ``GitHub Copilot CLI and Agent mode in Copilot Chat in IDEs
do not support content exclusion'' \cite{githubcopilotexclusion}. Cursor states that
``the terminal and MCP server tools used by Agent cannot block access to code
governed by \texttt{.cursorignore}'' \cite{cursorignore}.

\section{What the Arrangement Costs to Run}

The rules name the right paths, but they act on a command string, which is a coarser unit than a read. The difference shows up in the denial record. We matched
our own session transcripts against the harness's denial string and deduplicated
by the identifier the harness attaches to each tool call. Hence one refused invocation counts once however many copies of a transcript hold it, while a
retried command counts again. The result is 64 denial events over seven weeks of ordinary research work, from the second week of June to the end of July 2026. In 54 of them the refused invocation was a composite, several commands chained in one line, of which only one held a guarded token. Therefore the batch is refused whole, and the commands around
the offending one, which no rule objected to, are denied with it. This is the same limitation that keeps the rules from stating the requirement, since a rule names a command rather than a read.

Two command shapes, shown here with paths and hosts removed, account for most of
the composites. The first is a transfer folded into a chain of ordinary work,
\texttt{cd <dir> \&\& rsync -av <host>:<path> . \&\& ls -l}, where the transfer
trips the rule that confines the agent to one remote directory, and the two
commands around it are refused with it. The second is a rebuild, \texttt{rm -rf
<build> \&\& cmake -S . -B <build> \&\& make -j4}, where the deletion trips the destructive-verb gate, which refuses the configure
and the compile along with it. Sorted
by a classification rubric with a fixed order of precedence, 28 of the 64 are confinement boundaries, 26 the
destructive-verb gate, five deliberate probes of a boundary, four a protected target, and one fits none of these headings.

The protected-target count is a lower bound, because the rubric assigns by
precedence and puts that heading last. No denial in the corpus blocked ordinary
work on the protected files. Most came from boundaries unrelated to them, and a
project whose protected files are edited every day would produce a record
unlike this one. The corpus is drawn by matching the string that names a refused command. As a result, every event in the corpus is a command-surface denial, and no refusal on the built-in file tools can enter it. The categories are not mutually exclusive, and none of
these counts supports a rate.

\section{Related Systems and Scope of the Findings}

Mechanisms that can express the access rule do exist, though none closes the output channel the build opens. Each asks for something an
ordinary user does not hold, whether a privileged policy loader for kernel type
enforcement, a rewritten language runtime for access control inside a running
program, or the harness itself. Deploying any of them requires control of the layer beneath the agent.

One 2026 system comes closest, and its source is public. It
enforces policy beneath the tool surfaces by hooking operating-system events
across the agent's whole process tree, and it holds both constructs the rule
needs. A lineage gate ``checks whether the subject descends from a particular
process,'' and ``a child domain inherits all parent rules and may add local
rules, labels, or gates, but cannot remove, disable, or weaken any inherited
rule'' \cite{actplane2026}. So the rule can be written in that system. When the
agent starts the compiler, however, the compiler inherits the agent's label,
because the tags propagate through fork and exec, the two system calls by which
one process starts another. The protected file is then denied to the compiler, and
the build fails. Making the build work means naming the compiler
in the rule as an exception, and the agent can then invoke that exception
itself and read what the compiler prints.

Other systems cover parts of it. They check where the agent acts or where its output leaves, rather than where a file is opened. CaMeL decides what may leave the
system by where the data came from \cite{camel2025}. PORTICO binds a
repository path to an access mode and a bounded justification episode
\cite{portico2026}, which is the nearest match to the rule among these systems. It models a
single consumer, whereas this case has a second reader, the compiler, that keeps
opening the bytes. We did not find these systems applied to a requirement whose enforcement point is the read itself.

The result has a vendor-specific part and a general part. The exact tool scopes, the rule syntax and the documented exclusions belong to
one vendor and change with its releases. The general part has three
statements: a) matching a command string does not mediate a file open; b) a
rule that names a path cannot say which program is reading; c) a permission
that a child process inherits cannot distinguish the build from whatever
started the build. These hold wherever an agent runs programs on a user's
behalf. They, rather than the cell values, are what a reader should take from
the table.

This article assesses what the mechanisms available to an ordinary user can
express and cover. It does not offer an arrangement that holds against a hostile agent. We run this arrangement today, aware of its usefulness and limits.
The permission rules are now generated from a single inventory of protected
targets, which closes the gap found in the deny list's revisions.
A hook records every command the agent submits.
We accept the diagnostics and have not closed the compiler route, so anyone
running an arrangement like this one should assume that the build's
own output, the diagnostics and the compiled artifacts alike, is one place where protected content can still leave.

\section*{Availability and Prior Publication}

The configuration whose revision history this article reports is public under an MIT
license at \url{https://github.com/ShoRoy/claude-docker-sandbox}. Its five states
are the commits \texttt{fa5a7e7}, \texttt{e2377e8}, \texttt{2e5eb03},
\texttt{5dfaf89} and \texttt{cb24713}, in that order. The classification script behind
Table~\ref{tab:coverage} runs against that configuration and is released with
this article, together with the scripts that derive the revision history and the
denial counts. The denial corpus is not releasable, since it records real
project paths and a real remote host. Only counts and redacted shapes are
therefore reported.

Two practitioner articles by the author, published on Level Up Coding as
\emph{Dockered Claude}
(\url{https://levelup.gitconnected.com/dockered-claude-giving-an-ai-agent-the-keys-to-my-code-without-giving-it-the-house-5e23a8f2a342})
and its sequel
(\url{https://levelup.gitconnected.com/dockered-claude-keeping-research-code-compilable-but-obscured-from-my-ai-agent-c807d9ff7eb9}),
describe the deployment this article analyzes. The sequel is the disclosure from
which the three source-(A) routes of Table~\ref{tab:coverage} are taken. Neither
contains any of the results reported here. The requirement
statement, the coverage table, the revision history, the denial corpus and the
composition claim appear here for the first time, and no figure from either
article is reused.

\section*{Acknowledgments}

The coding agent this article studies, Claude Code (Anthropic), was also used
in producing it, under the arrangement described, for the literature search, the
released scripts and the figure, and as a writing aid throughout. The author
checked every citation against its source, revised every paragraph, and is
responsible for the content. The denial corpus was drawn from the
author's session transcripts and so includes the sessions on this manuscript up
to the end of July 2026.

\bibliographystyle{IEEEtran}
\bibliography{references}

\begin{IEEEbiographynophoto}{Shobhan Roy}
is a postdoctoral research scholar at IIHR--Hydroscience \& Engineering, University of Iowa, Iowa City, IA 52242, USA. His research interests include computational multiphysics, high-performance computing, and machine-learning surrogates for simulation. Roy received his PhD in mechanical engineering from the University of Iowa. He is a member of ACM SIGHPC. Contact him at shobhan-roy@uiowa.edu.
\end{IEEEbiographynophoto}

\end{document}